# Symmetry-Breaking Phase Transition without Peierls Mechanism in Conducting Monoatomic Chains

C. Blumenstein[1], J. Schäfer[1], M. Morresi[1], S. Mietke[2], R. Matzdorf[2], and R. Claessen[1]

[1]Physikalisches Institut, Universität Würzburg, 97074 Würzburg, Germany
[2]Fachbereich Naturwissenschaften, Universität Kassel, 34132 Kassel, Germany

The one-dimensional (1D) model system Au/Ge(001), consisting of linear chains of single atoms on a surface, is scrutinized for lattice instabilities predicted in the Peierls paradigm. By scanning tunneling microscopy and electron diffraction we reveal a second-order phase transition at 585 K. It leads to charge ordering with transversal and vertical displacements and complex interchain correlations. However, the structural phase transition is not accompanied by the electronic signatures of a charge density wave, thus precluding a Peierls instability as origin. Instead, this symmetry-breaking transition exhibits three-dimensional critical behavior. This reflects a dichotomy between the decoupled 1D electron system and the structural elements that interact via the substrate. Such substrate-mediated coupling between the wires thus appears to have been underestimated also in related chain systems.

Synthesis of atomic nanowires for electrical or mechanical applications may be achieved via self-organization of metal atoms on a semiconductor surface, yielding flat-lying chains of only a few atoms in width and of large extent [1]. At this scale, however, such one-dimensional (1D) systems are affected by structural and electronic instabilities. A prominent model for a transition in 1D is the Peierls instability [2]. It assumes a chain of atoms, as in Fig. 1(a), with a partially filled band, i.e. a metallic state. Peierls proposed that at low temperature (LT) such chain is instable against a *charge density wave (CDW)* [2], which arises generally from a nesting condition in the Fermi surface. This drives a metal-insulator transition (MIT) with energy gaps at the superstructure zone boundaries, accompanied by a periodic lattice distortion (PLD), see Fig 1(b) for a scenario with half-filling. Arguably, this picture is highly simplistic, because it neglects realistic interactions in a many-body solid such as multiple electron bands and collective lattice and orbital relaxation effects.

A recent theoretical study of Johannes and Mazin examined the stability of the Peierls phase [3]. Their analysis reveals that a pure CDW scenario is rather sensitive to deviations from the idealized 1D regime at finite temperature, imperfect nesting or scattering. It is hence concluded that a Peierls instability as the only driving mechanism for a PLD is highly unlikely to be observed in real quasi-1D materials.

Nonetheless, there have been several claims of experimental observations of a Peierls mechanism. Self-organized In nanowires on Si(111) render the first paradigmatic surface system where a nesting condition has been identified [4]. This seemingly matches the observed twofold PLD periodicity accompanied by a MIT upon moderate cooling to ~150 K. Moreover, the phase transition is masked by fluctuations in a wide temperature range above and below $T_C$ [5].

Yet, this Peierls interpretation is currently under heated debate. Subsequent theoretical studies argue with a competing model for In/Si(111), including a soft shear phonon responsible for the LT phase, which also produces an energy gap [6,7]. Furthermore, most recent calculations claim the transition to be entropy driven [8]. Likewise, the superstructure formation in Au/Si(557) chains [9,10] was previously discussed in terms of a CDW, yet is alternatively explained by buckling of step edge Si atoms [11,12]. Hence, the driving forces remain a fun-

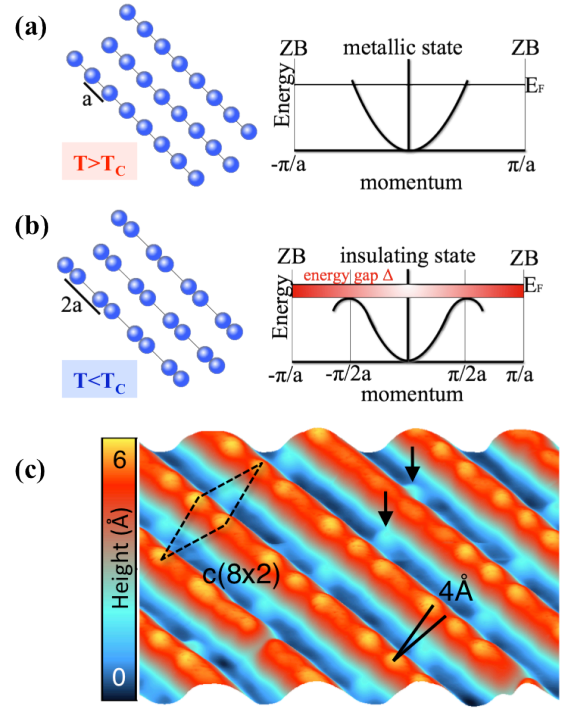

FIG. 1 (color online) (a) Above $T_C$: Atoms with one electron form a chain, leading to a metallic half-filled band. (b) Below $T_C$: a lattice instability with a dimerized PLD sets in, driven by nesting. Energy gaps at the new zone boundary lead to an insulating state. (c) STM image at 77 K of unoccupied states in Au/Ge(001) atom chains (+0.8 V, 0.4 nA, 10 × 6 nm). Triplets are identified, forming a fourfold superstructure relative to the high-T phase. Unit cell of high-T c(8×2) phase as dashed line. Arrows indicate protrusions in the groove.

damental issue: What is the contribution of a Peierls scenario in real 1D system phase transitions?

A recent addition to the class of surface defined 1D systems are atomic nanowires on the Ge(001) surface, formed by Pt [13,14,15] and Au [16,17]. Specifically the Au nanowires have a highly pronounced 1D architecture, see Fig. 1(c), since they are elevated above the substrate, separated by several lattice constants, and are of *atomic dimension* in width [17]. The Au/Ge(001) wires are solidly metallic at room temperature, rendering them a close representative of the simple atomic chain described by Peierls.

Studies of the electron system by means of angle-resolved photoelectron spectroscopy (ARPES) revealed a perfectly straight 1D Fermi surface, with several nesting conditions [18]. At low temperatures these chains even show a power-law suppression of the density of states at the Fermi level, which is characteristic for Luttinger liquid behavior [19]. Such physics may only occur in a nearly perfect 1D electron system, since coupling to higher dimensions would destroy this exotic state [20].

In this Letter, we scrutinize the 1D model system Au/Ge(001) over a wide temperature range, using scanning tunneling microscopy (STM) and low-energy electron diffraction (LEED). A second-order phase transition is found to occur at $T_C$ = 585 K, far above room temperature. Detailed investigation of structure and electronics does not show the characteristics of a Peierls mechanism, such as an energy gap, nesting or fluctuations. Instead, the determined exponent of $\beta \sim 1/3$ points at a 3D symmetry-breaking phase transition, which appears to be driven by the substrate. This questions the relevance of Peierls physics in such 1D chains at surfaces in general and has bearing for the long lasting dispute in related systems.

Experimentally, n-doped Ge(001) substrates were chemically etched and flashed to 1200 K in ultrahigh vacuum to produce a clean surface [21]. Subsequently, ~0.7 ML of Au was deposited with an electron beam evaporator onto the substrate held at ~500 °C. STM was conducted with an Omicron LT and VT apparatus.

Here we first explain he LT ground state of the Au/Ge(001) nanowires, including the long-range superstructure. The high structural order is apparent from the STM image of Fig. 1(c). The chains are equally spaced by 16 Å, and along the chains individual charge clouds are repeated every 8 Å, corresponding to a c(8×2) basic structure easily seen for most STM and LEED settings [16]. However, a complex fine structure exists in addition. For certain conditions, e.g. at a bias of +0.8 V as in Fig. 1(c), three spherical charge clouds appear elevated, giving rise to the *triplet* appearance. These *triplets* are spaced by 32 Å and have a lateral correlation such that they are shifted by 4 Å, e.g. in "up" direction in Fig. 1(c). This may be denoted in a *superstructure matrix*

$$\mathbf{M}_S = \begin{pmatrix} 0 & -8 \\ 4 & 1 \end{pmatrix}$$

Importantly, with equal probability we also observe a "down-shift" phase correlation of the triplets. A given interchain phase correlation (up or down) persists typically over five to seven wires until it reverses sign (see supplemental material) suggesting that both alignments are energetically equivalent. Since the spacing of the chains is very large, this lateral order already points at a weak coupling of the wire structure via the substrate.

Electron diffraction is also sensitive to this additional superstructure. The LEED pattern Fig. 2(a) includes both Ge(001)-terrace orientations, as well as the "up" and "down" interchain correlations. This diffraction pattern can be well reproduced (see supplemental material), by inclusion of the $\mathbf{M}_s$ superstructure on top of a c(8×2) reconstruction.

The charge landscape in STM changes drastically when detecting the orbitals of the *occupied* states in Fig. 2(b). A zigzag shape becomes visible. Importantly, these displacements are oriented *transversal* to the chain. The pattern is not a simple lateral corrugation, but consists of alternating "V"- and "W"-shaped segments, as also noted in [22]. Both *triplets* (unoccupied states) and *zigzag* (occupied states) share a periodicity of 32 Å. This may easily be seen by comparison of the line profiles at identical locations for both bias polarities, Figs. 2(c) and (d). The orbitals correlate such that the *V-shape* lies at the very same position as the *triplets* in the unoccupied

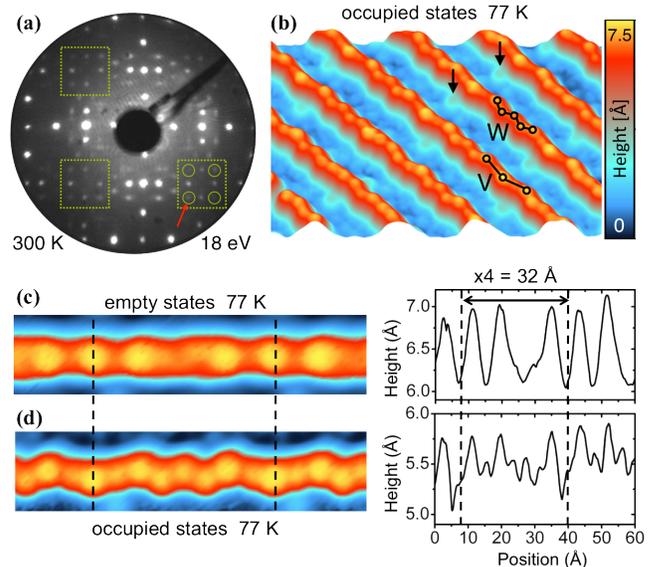

FIG. 2 (color online) (a) Dual-domain LEED pattern (18 eV) at 300 K, showing the basic c(8×2) structure and additional superstructure reflections. Red arrow indicates reflections for T-dependent analysis. (b) *Occupied* states STM image at 77 K (-0.8 V), same scale as Fig. 1(c). (c) Unoccupied states at 77 K (+0.8 V), and corresponding line profile. (d) Occupied states at 77 K (-0.8 V) of identical sample location. A zigzag structure appears between triplet sites. The long-range order contains W- and V-shaped segments.

intensity upon cooling. Effects from sample charging or decomposition of the structure can thereby be excluded. A small contribution to the intensity change resulting from the Debye-Waller effect can be estimated to be of the order of 2 % over the temperature range of 300 K, see supplemental material.

In thus interpreting the data as a continuous *second-order phase transition*, the measured superstructure intensity I(T) may be fitted with a power-law upon temperature [25] using

$$I(T) \propto \rho^2(T) \propto \left(\frac{T_C - T}{T_C}\right)^{2\beta} \qquad (1)$$

with the order parameter $\rho(T)$ which serves to minimize the free energy. It will be a measure of the transversal displacement $\xi(T)$ of the local charge density seen in STM. In diffraction experiments, the intensity I(T) of the superstructure Bragg scattering is proportional to $\rho^2(T)$ [23]. A close fit to the data is achieved for a critical temperature $T_C = 585 \pm 10$ K, and $\beta = 0.29 \pm 0.04$.

The disappearance of the transversal undulation is also seen in real space in the STM data of Fig. 3(b). Below $T_C$ the chains exhibit the characteristic 32 Å V-W zigzag. Above $T_C$ they only show marginal indications of a zigzag with 8 Å period, corresponding well to the c(8×2) structure. Furthermore, above $T_C$ no protrusions can be

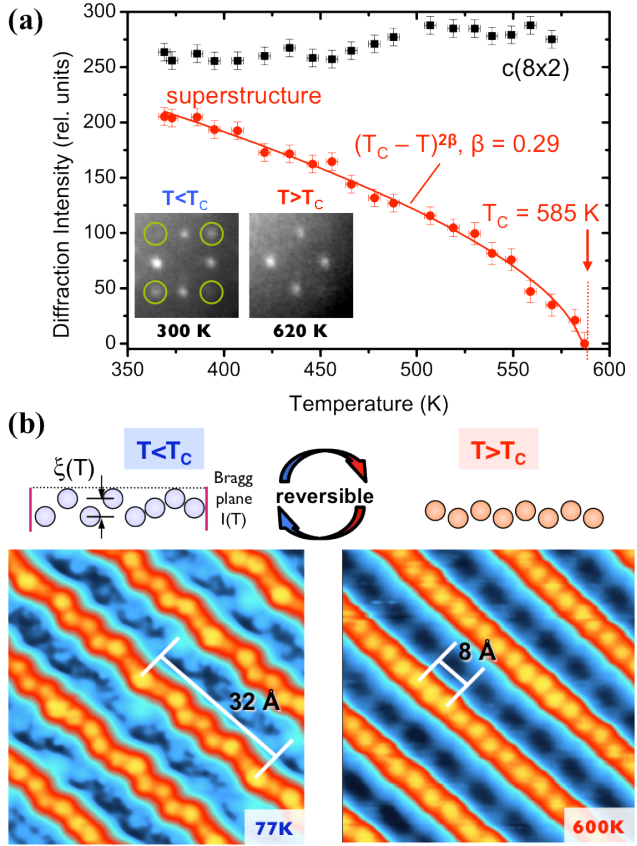

FIG. 3 (color online) (a) T-dependent LEED intensity analysis for basic c(8×2) (black) and superstructure (red) diffraction spots. The superstructure undergoes a 2nd-order phase transition with $T_C = 585$ K, while the underlying c(8×2) structure remains unaffected. Inset: Enlarged LEED images above and below $T_C$ (b) STM of occupied states (-0.8 V): at 77 K (left) the chains exhibit a characteristic V-W zigzag. At 600 K and above (right), the zigzag is replaced by a 8 Å period with minimal transversal buckling, thus matching the c(8×2) symmetry.

states.

Importantly, structure can even be resolved *between* the nanowires in Figs. 1(c) and Fig. 2(b). Both occupied and unoccupied states show small vertical protrusions in the groves close to the triplet locations, which are positioned closer to one side of the groove. The protrusions, spaced 32 Å along the chains, reflect the $\mathbf{M}_s$ superstructure as well. This is a further indication for the substrate to play a major role in mediating the charge correlations between the nanowires.

In looking for a possible phase transition, the Bragg reflection intensities of the superstructure were monitored using LEED. For the analysis, the superstructure reflections indicated by green circles in Fig. 2(a) have been chosen. We find a dramatic decrease of their intensity with increasing temperature, see Fig 3(a). Above 585 K the superstructure vanishes completely, while the c(8×2) reflections remain virtually unaffected by the temperature change, see inset to Fig. 3(a). This is a *reversible* process with full recovery of the superstructure

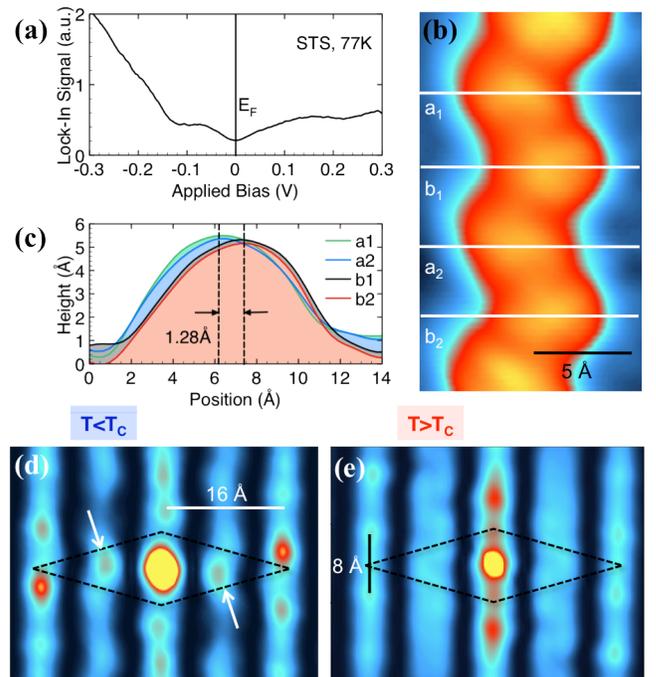

FIG. 4 (color online) (a) STS data, acquired with a lock-in amplifier at 77 K (50 spectra average), showing robust metallic behavior. (b) High resolution STM image of the zigzag chain element (-0.7 V, 0.4 nA, 1.1 × 2.4 nm) with four line profiles. (c) Line profile analysis shows a zigzag amplitude of 1.28 Å, indicating a single top atom. (d) and (e) Autocorrelated STM images (performed in k-space) below and above the phase transition. Between the chains additional protrusions appear in the troughs below $T_C$, (white arrows).

detected between the wires, which is evidence for the substrate to be involved in the transition as well.

The finding of a phase transition in such 1D system consequently imposes the question for the underlying driving forces. Does a *Peierls mechanism* play a vital role here? As is known from photoemission experiments [17, 18], the surface electron system of the nanowires consists of a single metallic band only, which has an extremely well defined 1D character. Hence a Peierls transition would induce an energy gap therein, leading to a MIT [2].

In contrast, the experimental tunneling spectrum (averaged over a unit cell at 77 K) in Fig. 4(a) gives proof of persistent metallicity in the ordered phase. The absence of a gap thus precludes a Peierls instability as origin of the observed symmetry-breaking phase transition. This conclusion is further corroborated by an analysis of the Fermi surface nesting situation [18]. While the Fermi surface indeed displays a clear 1D topology with three possible nesting vectors, the corresponding real space periodicities of $\lambda_1 = 66$ Å, $\lambda_2 = 28$ Å, and $\lambda_3 = 15$ Å clearly deviate from the observed 32 Å PLD repeat length.

Since a Peierls contribution can be excluded, a more detailed examination of the local effects of the phase transition on the structure has been performed, Fig 4(b). The W-shape of the LT phase has a transversal undulation amplitude $\xi$ of only 1.28 ± 0.03 Å at 77 K, as shown in the respective line profiles of Fig. 4(c). Upon heating this undulation is almost reduced to zero. Such transversal undulation is significantly too small to originate from a buckled dimer, since the buckling of a Ge-Ge dimer of clean Ge(001) amounts to 2.45 Å [24], which is almost twice as wide. Instead, assignment to a *single* atom on top is far more likely. While a Ge dimer on top is not supported by our results, the current findings rather point at a displacive character of the phase transition.

In order to scrutinize the periodic changes due to the phase transition, the autocorrelation of STM images below and above $T_C$ is shown in Fig. 4(d) and (e). Both images show the c(8×2) symmetry, the wire spacing and the 8 Å periodicity in chain direction. Clearly, in the LT image Fig. 4(d), additional intensity emerges at distances halfway between two wires. This intensity is identified with the protrusions observed in the grooves, Fig. 1(c). Therefore the phase transition also involves a movement *perpendicular* to the surface.

The dimensionality of the transition can be derived from the observed critical exponent β = 0.29. Within error bars our experimentally determined exponent agrees well with that of the *3D Ising model*, β = 0.33 [25]. Thus far, the only metal-semiconductor adsorbate system with an exponent in this range is (√3×√3)-Ag/Si(111), with β ~ 0.27 in diffraction studies [26,27]. That phase transition is discussed as an *order-disorder transition* with vibronic atom displacements in a 3D embedding. Interestingly, although this is a 2D reconstruction rather than a chain system, the substrate seems to mediate 3D interactions in both cases.

Further information on the dimensional character of the phase transition can be inferred from the temperature-evolution of the order parameter. While in an idealized 1D case quantum fluctuations hinder formation of a CDW condensate at any finite temperature, weak coupling to higher dimensions can stabilize the Peierls state at T > 0. Nonetheless, strong order parameter fluctuations will still persist above the transition up to the mean-field transition temperature [2]. In contrast, we observe a rather sharp transition at $T_C$, *not* masked by noticeable fluctuations. This also points to a *higher-dimensional character* of the structural instability.

The dichotomy between a 3D structural phase transition in a nanostructure which clearly shows 1D behavior of the conduction electrons [18,19] seems contradictory at first. However, for the total energy of the system also lower-lying states have to be taken into account, including the substrate backbonds as well as substrate-mediated bonding between the chains. One has to infer that the 1D electron system in the Au wires is completely decoupled from the underlying structural elements, which are involved in the phase transition. This also explains why a Peierls description is inapplicable to such systems: In the simple Peierls picture there is only one metallic electron band in the system, whereas in the Au/Ge(001) chains and in all other real world nanowires, many electrons from different orbitals contribute to the ground state.

Regarding these related nanowires with supposed electronically driven phase transitions, also alternative structural models were presented, e.g., for the low-T phase of In/Si(111) (hexagon model) [6,7] and likewise for the Au/Si(557) chains (step edge buckling) [11,12]. For the In chains remaining discrepancies to the experimental data from ARPES exist, since the electronic band structure is not completely reproduced by the hexagon model [28]. However, optical transitions are described rather well. Also, such modeling by density functional theory for principal reasons cannot describe the temperature-dependent phase transition *per se*, where entropy seems to play a vital role [8]. Nonetheless, our current findings strongly encourage that non-CDW-type transitions need to be considered more closely also for those nanowire systems.

In conclusion, the Au/Ge(001) atom chains exhibit a novel phase transition, which contrary to expectations is of 3D Ising type, and is most likely influenced by the substrate. The resulting interchain correlations involve a complex *lateral* and *vertical* structural rearrangement. This calls for a many-body view with coupled electronic and lattice degrees of freedom including the substrate, which also appear to be relevant for other low-dimensional nanostructures.

The authors acknowledge discussions with F. Assaad and M. Hohenadler. This work was supported by the DFG (grant Scha1510/2-1 and FOR1162).